%% file: main.tex
\documentclass[runningheads]{llncs}
\usepackage[T1]{fontenc}
\input{math_commands.tex}
\usepackage{hyperref}
\usepackage{mathtools}
\usepackage{graphicx}
\usepackage{booktabs}
\usepackage{multirow}
\usepackage{color}
\usepackage{pifont}
\usepackage{cite}
\usepackage[misc]{ifsym}

\definecolor{trolleygrey}{rgb}{0.5, 0.5, 0.5}
\definecolor{softgray}{RGB}{200, 200, 200}

\newcommand{\seg}[1]{\vspace{2mm} \noindent \textbf{#1}~}
\newcommand{\cmark}{\text{\ding{51}}}

\newcommand{\grayxmark}{\textcolor{softgray}{\ding{55}}}

\begin{document}
\title{UnWave-Net: Unrolled Wavelet Network for Compton Tomography Image Reconstruction}

\author{Ishak Ayad\inst{1,2}$^{(\textrm{\Letter},\thanks{Corresponding author})}$ \and
C\'ecilia Tarpau\inst{3} \and
Javier Cebeiro\inst{4} \and
Ma\"{\i} K. Nguyen\inst{1}
}

\authorrunning{I. Ayad et al.}
\titlerunning{Unrolled Wavelet Network for Compton Tomography Image Reconstruction}

\institute{
ETIS (UMR 8051), CY Cergy Paris University, ENSEA, CNRS, France \and
AGM (UMR 8088), CY Cergy Paris University, CNRS, France \and
Maxwell Institute for Mathematical Sciences \& School of Mathematical and Computer Sciences, Heriot-Watt University, United Kingdom \and
ITECA, UNSAM-CONICET, ECyT, Centro de Matemática Aplicada (CEDEMA), Universidad Nacional de San Mart\'{\i}n, Argentina\\
\email{ishak.ayad@cyu.fr}
}

\maketitle              
\input{sec/0_abstract}    
\input{sec/1_intro}
\input{sec/2_NCCCST}
\input{sec/3_methodology}
\input{sec/4_experiments}
\input{sec/5_conclusion}

\section{Acknowledgments}
This work was granted access to the HPC resources of IDRIS under the allocation 2021-[AD011012741] / 2022-[AD011013915] provided by GENCI and supported by DIM Math Innov funding.

\bibliographystyle{splncs04}
\bibliography{biblio}

\end{document}

%% file: math_commands.tex
\usepackage{amsmath,amsfonts,bm}

\newcommand{\regfunc}[1]{{\mathcal #1}}

\def\eqref#1{equation~\ref{#1}}

\def\1{\bm{1}}

\def\vf{{\bm{f}}}

\def\vw{{\bm{w}}}
\def\vx{{\bm{x}}}
\def\vy{{\bm{y}}}
\def\vz{{\bm{z}}}

\def\mA{{\bm{A}}}

\DeclareMathAlphabet{\mathsfit}{\encodingdefault}{\sfdefault}{m}{sl}
\SetMathAlphabet{\mathsfit}{bold}{\encodingdefault}{\sfdefault}{bx}{n}

\def\gA{{\mathcal{A}}}

\def\gK{{\mathcal{K}}}

\def\gR{{\mathcal{R}}}

\newcommand{\R}{\mathbb{R}}

\newcommand{\normltwo}{L^2}

\newcommand{\argmin}[1]{\underset{#1}{\operatorname{arg}\,\operatorname{min}}\;}

\newcommand\norm[1]{\left\lVert#1\right\rVert}

%% file: sec/0_abstract.tex
\begin{abstract}
Computed tomography (CT) is a widely used medical imaging technique to scan internal structures of a body, typically involving collimation and mechanical rotation. Compton scatter tomography (CST) presents an interesting alternative to conventional CT by leveraging Compton physics instead of collimation to gather information from multiple directions.
While CST introduces new imaging opportunities with several advantages such as high sensitivity, compactness, and entirely fixed systems, image reconstruction remains an open problem due to the mathematical challenges of CST modeling. In contrast, deep unrolling networks have demonstrated potential in CT image reconstruction, despite their computationally intensive nature.
In this study, we investigate the efficiency of unrolling networks for CST image reconstruction. To address the important computational cost required for training, we propose UnWave-Net, a novel unrolled wavelet-based reconstruction network.
This architecture includes a non-local regularization term based on wavelets, which captures long-range dependencies within images and emphasizes the multi-scale components of the wavelet transform.
We evaluate our approach using a CST of circular geometry which stays completely static during data acquisition, where UnWave-Net facilitates image reconstruction in the absence of a specific reconstruction formula.
Our method outperforms existing approaches and achieves state-of-the-art performance in terms of SSIM and PSNR, and offers an improved computational efficiency compared to traditional unrolling networks.
\keywords{Compton Scatter Tomography \and Image reconstruction \and Unrolling networks \and Wavelet transform.}
\end{abstract}

%% file: sec/1_intro.tex
\section{Introduction and Related Works}
Computed tomography (CT) is widely used in clinical practice for scanning internal body structures. These devices use collimation to obtain directional x-ray information and mechanical rotation of the source-detector pair around the object to gather complete sets of data. Compton scatter tomography (CST) is an alternative to CT, utilizing Compton physics instead of collimation to obtain directional information of transmitted photons. The directional shift $\omega$ observed in Compton scattering correlates with the energy loss of the photon, transitioning from its initial energy $E_0$ to $E(\omega)$, according to the following one-to-one relation:

\begin{equation*}
\label{eq:compton}
E(\omega)=\frac{E_0}{1+ \frac{E_0}{mc^2}(1-\cos \omega)},
\end{equation*}

\noindent where $mc^2 = 0.511$ MeV. CST introduces novel possibilities for imaging (assuming the use of energy-resolved sensors), particularly in terms of designing new geometries; for example, compact and/or completely fixed systems can now be contemplated~\cite{Norton2019,TN2011,TCMN2019,tarpau2020compton,TCNRM2020,cebeiro2021ip}. In contrast to traditional CT, certain devices offer scanning configurations with the object positioned outside the system~\cite{Webber_2020,cebeiro2021ip,tarpau2020compton,INTECH}. However, CST introduces new challenges in image reconstruction due to data acquisition, which involves the use of generalized Radon transforms on families of circular arcs in 2D. Present literature on this subject is confined to theoretical results providing analytical reconstruction formulas for a limited CST geometries. 

With deep learning's success in CT~\cite{fbpconvnet,ddnet,dudotrans,gloredi} and MRI~\cite{endtoend,humusnet} reconstruction, two main categories emerged: Post-processing methods like FBPConvNet~\cite{fbpconvnet} and DuDoTrans~\cite{dudotrans} treat reconstruction as a denoising task. While effective in addressing artifacts, they often struggle with global information recovery from sparse data. Conversely, unrolling networks have emerged as an enticing technique for image reconstruction~\cite{learnedpd,cnngrad,learn,admmct,dior,fistanet,regformer}, treating it as an optimization task resulting in an iterative algorithm like gradient descent, and subsequently unrolled into a neural network to learn the regularization terms. However, unrolled networks face two primary challenges: firstly, the difficulty in capturing long-range dependencies due to reliance on locally-focused regularization terms using CNNs; secondly, unrolled networks consist of a cascade of network layers to mimic the iterative reconstruction process, leading to high computational costs.

In this paper, we explore unrolling networks~\cite{aun} efficiency for CST image reconstruction. Furthermore, to mitigate memory and computational costs of unrolling networks, we introduce UnWave-Net, a novel deep unrolled network incorporating a wavelet-based regularization term. Our approach employs a discrete wavelet transform on the input features~\cite{wdpm,wnerf}, decomposing them into low-frequency (LL) and high-frequency (LH, HL, HH) components across four sub-bands. The regularization term is specifically applied to the low-frequency features, resulting in a significant reduction in learning and inference times while preserving quality. To demonstrate the efficacy of UnWave-Net, we consider a non-collimated circular CST (NCCCST) system, comprised of a source and sensors positioned on a circular ring. The analytic inversion of the corresponding Radon transform remains an open problem. We illustrate that the UnWave-Net represents a compelling method for image reconstruction in NCCCST, achieving state-of-the-art results in both training and inference speed, along with superior performance in quantitative metrics. The main contributions of this paper are:

\begin{enumerate}
    \item UnWave-Net, a wavelet unrolled network for CST image reconstruction leveraging wavelet subbands' dimensional reduction to expedite inference.
    \item Achievement of state-of-the-art results in training and inference times, along with superior quantitative performance through experiments conducted on a CST modality, which lacks an inverse analytic formula for image reconstruction.
\end{enumerate}

%% file: sec/2_NCCCST.tex
\section{Modelling of the Non-Collimated Circular CST}
\label{sec:NCCCST}

\begin{figure}[htbp]
\centering
\includegraphics[width=\textwidth]{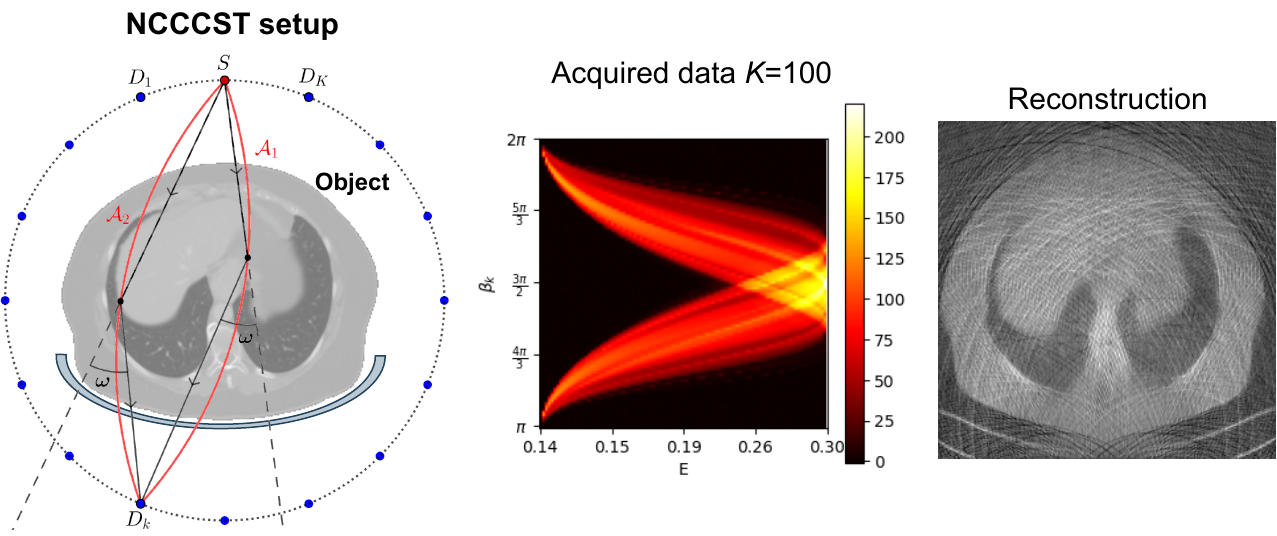}
\caption{NCCCST setup. Left: the dotted circle represents the ring containing point-like blue detectors $D_{k}$, with the red source $S$. Center: measured data. Right: reconstruction using pseudo-inversion of the data exhibits severe artifacts.}
\label{fig:ccst} 
\end{figure}

The NCCCST system consists of a source $S$ and $K$ fixed uncollimated detectors $D_{k, k\in\{1,\dots K\}}$ placed on a radius of $P/2$ including the source. Detectors are non-collimated and energy-resolved. The scanned object is within the ring, and the source has a plate collimator confining photons to a plane, defining, thus, a 2D scanning slice (see Fig.~\ref{fig:ccst}). NCCCST inherits advantages from its predecessor CCST~\cite{TCNRM2020, tarpau2020compton}, such as non-moving components and a compact design. First-order scattered photons recovered by a detector $D_{k}$ at the same energy $E(\omega)$ correspond to scattering sites on one of two circle arcs with $S$ and $D_{k}$ as endpoints, subtending the angle $(\pi-\omega)$. An example of such a pair of scanning arcs is depicted in Fig.~\ref{fig:ccst}. The data model considers only first-order scattering, with higher orders treated as noise. Attenuation effects are neglected. In contrast to CCST~\cite{TCNRM2020}, NCCCST doesn't determine if the scattering site lies on circle arc $\gA_1$ or $\gA_2$ given a scattering angle $\omega$.

In the following calculations, the source $S$ is placed at the origin of the coordinate system. Let $f$ be a non-negative function with compact support inside a disc of diameter $P$ centred at $O = (0, -P/2)$. The Radon-type model mapping $f$ to its integrals over the family of double circular arcs $\gA_1\cup\gA_2$ is given by:
\begin{align}
    \begin{split}
        \gR f(\omega, \beta_k) = \!\smashoperator[r]{\int\limits_{-\infty}^{\infty}} \Bigg[ \smashoperator[r]{\int\limits_{\beta_k}^{\beta_k+\omega}} f(r,\theta)\gK(\omega,\beta_k;r,\theta) d\theta + \!\!\!\!\!\! \smashoperator[r]{\int\limits_{\beta_k-\omega}^{\beta_k}} f(r,\theta)\gK(-\omega,\beta_k;r,\theta) d\theta \Bigg] dr,
    \label{eq:cirarcRT_wo_coll}
    \end{split}
\end{align}
\noindent where $\gK(\omega,\beta_k;r,\theta) = \rho(\omega, \beta_k) \delta(r-\rho(\omega, \beta_k)\cos(\theta-\phi(\omega, \beta_k)))$ is a delta kernel defining the scanning circle arcs with diameter $\rho(\beta_k, \omega) = P\cos\left(\beta_k+\frac{\pi}{2}\right)/\sin{(\omega)}$ and relative angle $\phi(\beta_k, \omega) =\beta_k + \omega -\pi/2$ to the $x$-axis. Variables are the scattering angle $\omega$, the angle subtended by the detector $\beta_k$, and the polar variables $(r,\theta)$. Function $f$ represents the electronic density of the object and $\gR f(\omega, \beta_k)$ is the data measured by detectors. An example of collected data is presented in Fig.~\ref{fig:ccst}. Currently, neither inverse formulas nor inversion results are known for the model on pairs of circles (Eq. \ref{eq:cirarcRT_wo_coll}). This is mainly due to the lack of shift or rotational invariance in the direct operator, motivating our work to find a solution for image reconstruction, i.e., the inversion of Eq. \ref{eq:cirarcRT_wo_coll}.

%% file: sec/3_methodology.tex
\section{Methodology}

\subsection{Related Theories}

\seg{Inverse Problem Formulation.}
NCCCST image reconstruction problem can be mathematically formalized as the solution to a linear equation in the form of:
\begin{align}
\vy=\mA\vx,
\label{eq:linear}
\end{align}
\noindent where $\vx \in \mathbb{R}^n$ represents the object to reconstruct ($n = h \times w$), and $\vy \in \mathbb{R}^m$ denotes the data ($m = K \times N_E$), where $K$ is the number of detectors and $N_E$ is the number of finite elements in the energy domain $[E(\pi), E_0]$. $\mA \in \mathbb{R}^{n \times m}$ is the discrete forward model. The objective is to recover the object $\vx$ from the observed data $\vy$. Since CST image reconstruction deals with an ill-posed inverse problem, pseudo-inversion is insufficient for qualitative reconstruction (see Fig.~\ref{fig:ccst}). Iterative reconstruction algorithms are employed to minimize a regularized objective function with an $\normltwo$ norm constraint:
\begin{align}
        \hat{\vx}=\argmin{\vx} J(\vx)=\frac{\lambda}{2} \norm{\mA\vx - \vy}^2_2 + \regfunc{F}(\vx),
        \label{eq:objective}
\end{align}
where $\regfunc{F}(\vx)$ is the regularization term. Initially, ill-posed problems were addressed using optimization techniques like the truncated singular value decomposition (SVD) algorithm~\cite{svd} or iterative approaches such as algebraic reconstruction technique (ART)~\cite{art}. Additionally, techniques like total variation~\cite{tv} and Tikhonov regularization~\cite{tikhonov} can improve the reconstructions.

\seg{Deep Unrolling Networks.}
Assuming that $\regfunc{F}$ is differentiable and convex, gradient descent can be used to solve Eq. \ref{eq:objective}:
\begin{equation}
    \vx_{t+1}=\vx_{t} - \alpha (\lambda \mA^{\dagger}\left(\mA\vx_t - \vy\right) + \nabla_{\vx} \regfunc{F}(\vx_t)),
\label{eq:gradient_solution}
\end{equation}
where $\alpha$ is the step size (neglected for redundancy with the learned parameters), and $\mA^{\dagger}$ is the pseudo-inverse. Optimization limitations, like manual parameter selection, are tackled by recent deep learning methods~\cite{regformer,learn}. By allowing terms in Eq. \ref{eq:gradient_solution} to be iteration dependent, the gradient descent iteration becomes:
\begin{align}
    \vx_{t+1}=\vx_{t} - \lambda_{t} \mA^{\dagger}\left(\mA\vx_{t} - \vy\right) + \regfunc{G}(\vx_{t}),
\label{eq:unrolled_solution}
\end{align}
where $\regfunc{G}$ is a learned mapping representing the gradient of the regularization term. Finally, Eq. \ref{eq:unrolled_solution} is unrolled into a deep recurrent neural network to learn the optimization parameters and the regularization term.

\begin{figure}[htbp]
\centering
\includegraphics[width=\textwidth]{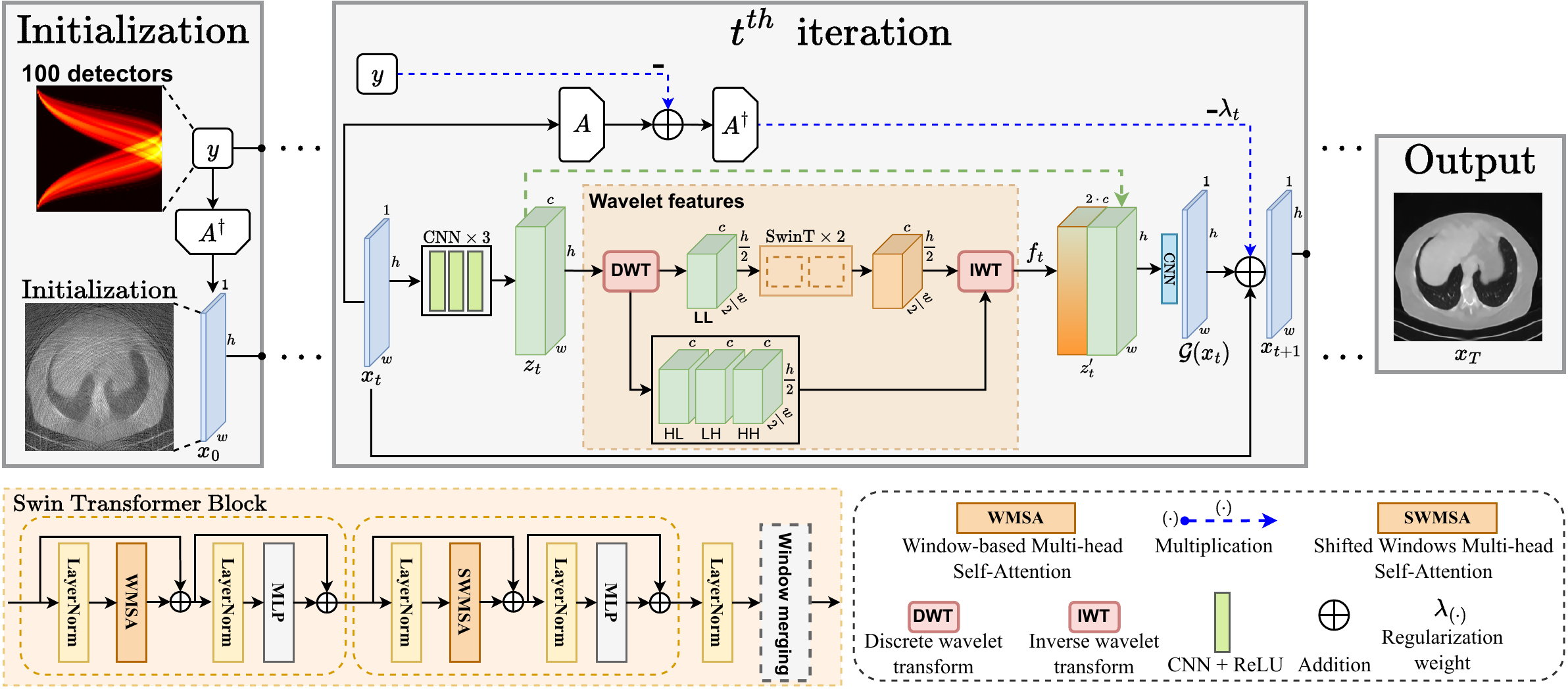}
\caption{Overall structure of the proposed UnWave-Net for NCCCST reconstruction. The method utilizes wavelet transform to reduce computational complexity.}
\label{fig:unwavenet} 
\end{figure}

\subsection{The proposed UnWave-Net}
Recent studies on unrolling networks have explored various representations of the regularization term gradient (denoted as $\regfunc{G}$ in Eq. \ref{eq:unrolled_solution}), from conv-nets~\cite{learn,learnpp,endtoend} to attention-based networks~\cite{regformer,humusnet}. As deep learning models grow in complexity, the representation of the regularization term gradient becomes more computationally and memory-intensive~\cite{aun}. To mitigate this, we introduce UnWave-Net, a novel unrolling network that leverages the wavelet transform to represent the regularization term gradient. Recently, the wavelet transform has been widely used to reduce the complexity of deep learning models, such as diffusion probabilistic models~\cite{wdpm}, or in NeRF models for generalizable, high-quality synthesis~\cite{wnerf}. Inspired by this, we apply a CNN layer to the input image $\vx_t \in \R^{h \times w \times 1}$ at iteration $t$ to obtain $\vz_t \in \R^{h \times w \times c}$ feature map, which is then transformed into wavelet coefficients $\vw_t \in \R^{\frac{h}{2} \times \frac{w}{2} \times 4 \cdot c}$. The LL-frequencies undergo Swin Transformer processing~\cite{swint} to capture long-range dependencies, while the HL, LH, and HH frequencies remain unchanged. After inverse wavelet transformation, the new feature map $\vf_t \in \R^{h \times w \times c}$ is concatenated with $\vz_t$ to yield final feature map $\vz'_t = \text{cat}(\vz_t, \vf_t) \in \R^{h \times w \times 2 \cdot c}$. Lastly, $\vz'_t$ passes through a CNN layer to produce the regularization term gradient $\regfunc{G}(\vx_t) \in \R^{h \times w \times 1}$. The overall architecture of the proposed UnWave-Net is depicted in Fig.~\ref{fig:unwavenet}. The network takes as inputs the acquired data $\vy$ and the corresponding pseudo-inverse reconstruction $\mA^{\dagger} \vy$. It consists of $T$ iterations, each incorporating a wavelet regularization term.

%% file: sec/4_experiments.tex
\begin{figure}[htpb]
  \centering
  \includegraphics[width=\textwidth]{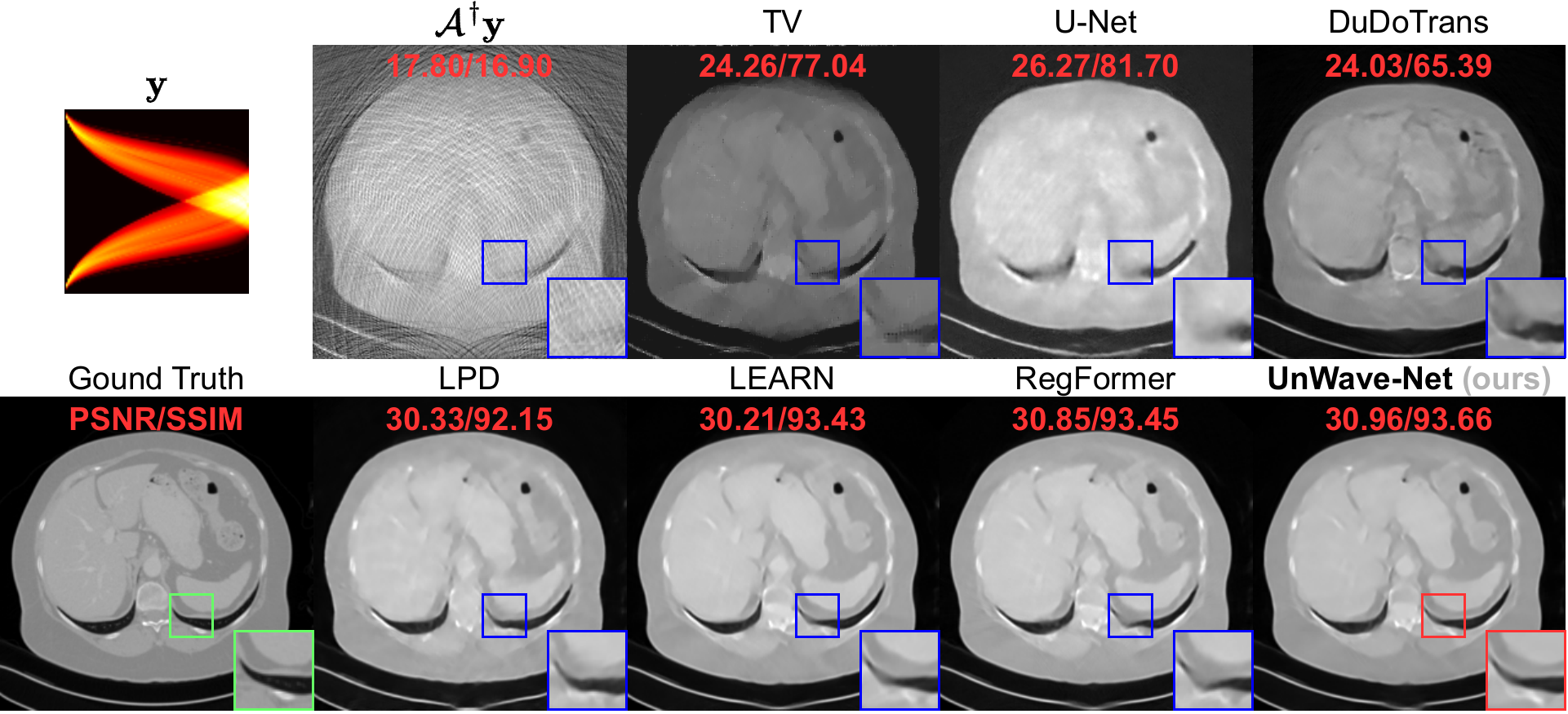}
  \caption{Visual comparison with noisy data and $K = 100$. The cropped region highlights the superior artifact supression of our UnWave-Net compared to RegFormer.}
  \label{fig:visual}
\end{figure}

\section{Experiments}
\subsection{Experimental Setup}
\seg{Dataset and Evaluation metrics.}
We evaluate our method on the AAPM dataset~\cite{aapm}, consisting of $2378$ full-dose CT images with 3mm thickness from $10$ patients. In line with standard practices~\cite{learn,dudotrans,regformer}, we employ peak signal-to-noise ratio (PSNR) and structural similarity index measure (SSIM) for quantitative evaluation.

\seg{Implementation and Training details.}
The AAPM dataset is divided into $1920$ training images from $8$ patients, $244$ validation images from $1$ patient, and $214$ test images from the remaining patient, all resized to $256 \times 256$. The considered NCCCST is made of $K \in \{100, 150\}$ detectors of energy resolution $\Delta E = 0.0016$ MeV. The source is assumed to be mono-energetic of energy $E_0 = 0.3$ MeV. The forward operator is simulated by discretizing Eq. \ref{eq:cirarcRT_wo_coll}. To mimic real-world CT scanning, mixed noise is applied to the acquired data, comprising $5\%$ Gaussian noise and Poisson noise with an intensity of $1 \times 10^6$. Model training involves $50$ epochs using $4$ Nvidia Tesla V100 GPUs (32GB RAM). The AdamW optimizer~\cite{adamw} with a learning rate of $1 \times 10^{-4}$ and weight decay $1 \times 10^{-2}$ is used, along with MSE loss and a batch size of $1$. A learning rate decay factor of $0.1$ is applied after $40$ epochs. The unrolling iterations for UnWave-Net are set to $T=16$, and a Haar wavelet transform with a feature map depth of $48$ is employed for regularization. The sizes of the convolution kernel are $5 \times 5$. For the SwinT model, a window size of $8$ and a depth of $2$ are utilized.

\seg{State-of-the-art baselines.}
We compare UnWave-Net with Chambolle-Pock's algorithm using a handcrafted total variation (TV) regularizer and several state-of-the-art CT reconstruction models: (1) post-processing methods, namely U-Net~\cite{fbpconvnet} and DuDoTrans~\cite{dudotrans}; (2) deep unrolling networks, including LPD~\cite{learnedpd}, LEARN~\cite{learn}, and RegFormer~\cite{regformer}. It is important to note that the FBP operators in these methods are replaced by pseudo-inverse reconstruction due to the lack of an analytical solution for the NCCCST problem. To ensure fair comparison, we use authors' code or meticulously implement methods based on their papers.

\input{tab/aapm_quantitative}

\subsection{Comparison with state-of-the-art methods}

\seg{Quantitative and Visual comparisons.}
We evaluate our model against state-of-the-art baselines employing two different numbers of detectors, $K \in \{100, 150\}$, and considering two levels of noise: noise-free and noisy data. The summarized results are presented in Table~\ref{tab:aapm_quantitative}. Our approach, UnWave-Net, consistently achieves state-of-the-art performance across all scenarios. It surpasses the second-best baseline, RegFormer, with an average improvement of ($+0.03\%$ SSIM, $+0.3$ dB PSNR) and ($+0.1\%$ SSIM, $+0.49$ dB PSNR) under noise-free and noisy conditions, respectively. Visual results are depicted in Fig.~\ref{fig:visual}, where U-Net, DuDoTrans, and LPD produce blurry images with noticeable artifacts. Although RegFormer and LEARN perform adequately, they lack detail in the lung and aorta regions. In contrast, UnWave-Net consistently generates detailed, high-quality images.

\seg{Efficiency comparison.}
The results in Table~\ref{tab:costs_aapm}, conducted on one V100 GPU, show that UnWave-Net is computationally more efficient than gradient descent-based unrolling networks, outperforming LEARN and RegFormer with speedups of $1.36\times$ and $1.31\times$, respectively. Moreover, UnWave-Net utilizes less memory than RegFormer, reducing memory usage by $1.13\times$, approaching the efficiency of LPD. Additionally, our method requires only $16$ unrolling iterations compared to $30$ for LEARN and $18$ for RegFormer, all while achieving superior performance.

\input{tab/costs_aapm}

\subsection{Ablation Study}
\begin{figure}[htpb]
  \centering
  \includegraphics[width=\textwidth]{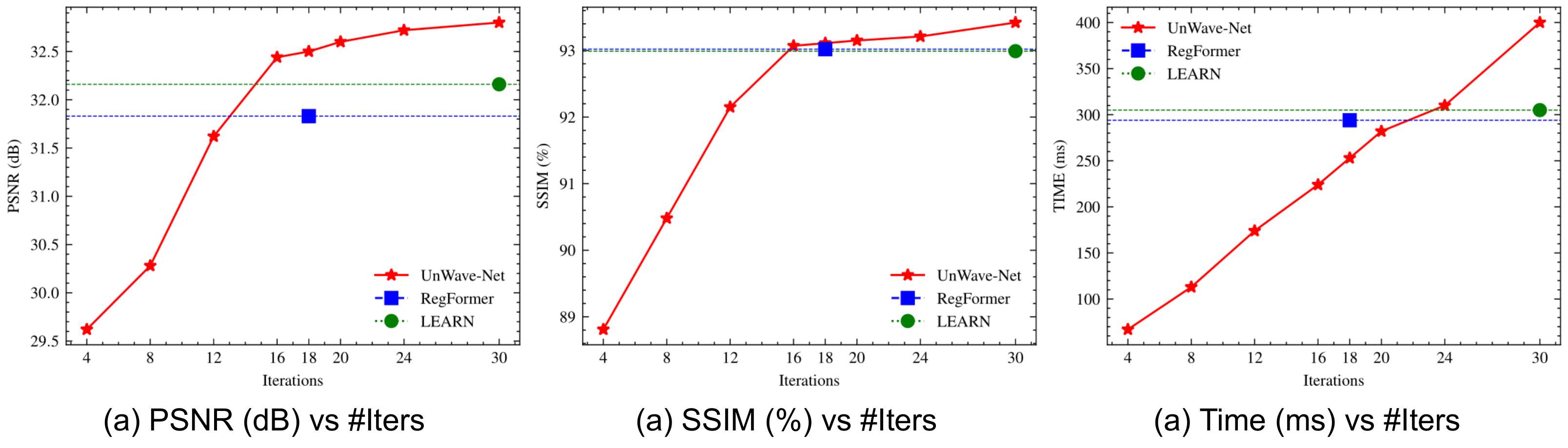}
  \caption{Ablation on the number of unrolling iterations. Quantitative and efficiency comparisons of our UnWave-Net against RegFormer and LEARN.}
  \label{fig:ablation}
\end{figure}

To assess how the number of unrolling iterations affects UnWave-Net's performance, we conducted an ablation study on the AAPM dataset with $K = 100$. Fig.~\ref{fig:ablation}a and Fig.~\ref{fig:ablation}b illustrate that even with $T \leq 16$, UnWave-Net achieves higher PSNR and SSIM metrics compared to RegFormer and LEARN. This underscores the effectiveness of UnWave-Net over other gradient descent-based unrolling networks. Additionally, UnWave-Net exhibits significantly lower time costs than RegFormer and LEARN, as shown in Fig.~\ref{fig:ablation}c, even when $T = 20$. This showcases the efficiency of our method in terms of inference time.

%% file: tab/aapm_quantitative.tex
\begin{table}[htpb]
\caption{Quantitative evaluation on AAPM of state-of-the-art methods (PSNR in dB and SSIM in \%). \textbf{Bold}: Best, \underline{under}: second best.}\label{tab:aapm_quantitative}
\resizebox{1.0\textwidth}{!}{
\begin{tabular}{lllll@{\hskip 15pt}llll}
\toprule
\multirow{3}{*}{\textbf{Method}} & \multicolumn{4}{c}{\textbf{No noise}} & \multicolumn{4}{c}{\textbf{With noise}} \\
\cmidrule(lr){2-5}
\cmidrule(lr){6-9}
{} & \multicolumn{2}{c}{\textbf{$K=150$}} & \multicolumn{2}{c}{\textbf{$K=100$}} & \multicolumn{2}{c}{\textbf{$K=150$}} & \multicolumn{2}{c}{\textbf{$K=100$}} \\
\cmidrule(lr){2-3}
\cmidrule(lr){4-5}
\cmidrule(lr){6-7}
\cmidrule(lr){8-9}
 & {PSNR $\uparrow$} & {SSIM $\uparrow$} & {PSNR $\uparrow$} & {SSIM $\uparrow$} & {PSNR $\uparrow$} & {SSIM $\uparrow$} & {PSNR $\uparrow$} & {SSIM $\uparrow$}\\
\midrule
Pseudo-inverse & 19.52 & 21.90 & 18.21 & 17.26 & 19.43 & 21.44 & 18.17 & 17.08\\

TV & 26.61 & 80.56 & 25.33 & 78.31 & 26.54 & 80.51 & 25.27 & 78.19\\

U-Net~\cite{fbpconvnet} & 28.32 & 83.59 & 27.14 & 81.47 & 28.03 & 83.93 & 27.36 & 82.58\\

DuDoTrans~\cite{dudotrans} & 29.16 & 81.96 & 27.32 & 80.08 & 28.99 & 81.46 & 27.36 & 79.74\\

LPD~\cite{learnedpd} & 32.07 & 94.05 & 31.48 & 92.86 & 31.43 & 92.72 & 31.40 & 91.83\\

LEARN~\cite{learn} & 31.92 & \underline{95.11} & 31.90 & 94.08 & 31.63 & 93.66 & \underline{32.16} & 92.95\\

RegFormer~\cite{regformer} & \underline{32.45} & 95.02 & \underline{32.54} & \textbf{94.22} & \underline{32.11} & \underline{93.80} & 31.83 & \underline{92.98}\\

\cmidrule(l){1-9}
UnWave-Net~\textcolor{trolleygrey}{(ours)} & \textbf{32.86} & \textbf{95.12} & \textbf{32.73} & \underline{94.18} & \textbf{32.48} & \textbf{93.91} & 
\textbf{32.44} & \textbf{93.07}\\

\bottomrule
\end{tabular}
}
\end{table}

%% file: tab/costs_aapm.tex
\begin{table}[htbp]
\caption{Efficiency comparison of state-of-the-art methods with $K=100$.}\label{tab:costs_aapm}
\setlength{\tabcolsep}{4pt}
\resizebox{1.0\textwidth}{!}{
\begin{tabular}{l c c c c c c}
\toprule
Method & Unrolled & \#Iters & Epoch time (s) & \#Params (M) & Memory (GB) & Inference time (ms)\\
\toprule
TV & \grayxmark & 500 & - & - & - & 5600\\
U-Net~\cite{fbpconvnet} & \grayxmark & - & 80 & 31.1 & 3.94 & 12.7\\
DuDoTrans~\cite{dudotrans} & \grayxmark & - & 115 & 15.0 & 3.80 & 27.5\\
LPD~\cite{learnedpd} & \cmark & 10 & 370 & 0.25 & 3.63 & 86.4\\
LEARN~\cite{learn} & \cmark & 30 & 1600 & 3.50 & 3.64 & 305\\
RegFormer~\cite{regformer} & \cmark & 18 & 1710 & 4.84 & 4.22 & 294\\
\cmidrule(l){1-7}
UnWave-Net~\textcolor{trolleygrey}{(ours)} & \cmark & 16 & 1270 & 2.80 & 3.72 & 224\\
\bottomrule
\end{tabular}
}
\end{table}

%% file: sec/5_conclusion.tex
\section{Conclusion}
In this paper, we present UnWave-Net, a wavelet-based deep unrolled network tailored for CST imaging. Through extensive experimentation on a stationary CST system with circular geometry (i.e., NCCCST), UnWave-Net demonstrates superior performance in both image quality and reconstruction speed, positioning it as a state-of-the-art solution in this domain. Notably, our study underscores the adaptability and effectiveness of established unrolling networks in addressing novel imaging challenges, showcasing their versatility and broad applicability. Moreover, we emphasize the practical advantage of UnWave-Net's robustness to noisy data, enhancing its feasibility in real-world scenarios. We envision UnWave-Net as a promising method for CST imaging with potential extensions to other imaging modalities. Nonetheless, it is essential to acknowledge that our approach inherits the inherent drawback of deep unrolled networks, particularly in terms of prolonged training time compared to post-processing methods. Future research efforts will focus on addressing these limitations.

%% file: main.bbl
\begin{thebibliography}{10}
\providecommand{\url}[1]{\texttt{#1}}
\providecommand{\urlprefix}{URL }

\bibitem{Norton2019}
Norton, S.J.: Compton-scattering tomography with one source and one detector: a simple derivation of the filtered-backprojection solution. Inverse Problems in Science and Engineering pp. 1--12 (2019)

\bibitem{TN2011}
Truong, T.T., Nguyen, M.K.: {R}adon transforms on generalized {C}ormack's curves and a new {C}ompton scatter tomography modality. Inverse Problems  \textbf{27} (2011)

\bibitem{TCMN2019}
Tarpau, C., Cebeiro, J., Morvidone, M.A., Nguyen, M.K.: A new concept of {C}ompton scattering tomography and the development of the corresponding circular {R}adon transform. IEEE TRPMS  \textbf{4},  433--440 (2020)

\bibitem{tarpau2020compton}
Tarpau, C., Nguyen, M.K.: Compton scattering imaging system with two scanning configurations. Journal of Electronic Imaging  \textbf{29} (2020)

\bibitem{TCNRM2020}
Tarpau, C., et~al.: Analytic inversion of a {R}adon transform on double circular arcs with applications in {C}ompton scattering tomography. IEEE TCI  \textbf{6},  958--967 (2020)

\bibitem{cebeiro2021ip}
Cebeiro, J., et~al.: On a three-dimensional {C}ompton scattering tomography system with fixed source. Inverse Problems  (2021)

\bibitem{Webber_2020}
Webber, J., Miller, E.L.: Compton scattering tomography in translational geometries. Inverse Problems  \textbf{36} (2020)

\bibitem{INTECH}
Truong, T.T., Nguyen, M.K.: Recent Developments on {C}ompton Scatter Tomography: Theory and Numerical Simulations. IntechOpen (2012)

\bibitem{fbpconvnet}
Jin, K.H., et~al.: Deep convolutional neural network for inverse problems in imaging. IEEE Transactions on Image Processing  \textbf{26},  4509--4522 (2017)

\bibitem{ddnet}
Zhang, Z., et~al.: A sparse-view {CT} reconstruction method based on combination of {DenseNet} and deconvolution. IEEE TMI  \textbf{37},  1407--1417 (2018)

\bibitem{dudotrans}
Wang, C., et~al.: {DuDoTrans}: Dual-domain transformer for sparse-view {CT} reconstruction. In: Machine Learning for Medical Image Reconstruction. pp. 84--94 (2022)

\bibitem{gloredi}
Li, Z., et~al.: Learning to distill global representation for sparse-view {CT}. In: ICCV. pp. 21196--21207 (2023)

\bibitem{endtoend}
Sriram, A., et~al.: {E}nd-to-{E}nd variational networks for accelerated {MRI} reconstruction. In: MICCAI. pp. 64--73 (2020)

\bibitem{humusnet}
Fabian, Z., Tinaz, B., Soltanolkotabi, M.: {HUMUS}-{N}et: Hybrid unrolled multi-scale network architecture for accelerated {MRI} reconstruction. In: NeurIPS (2022)

\bibitem{learnedpd}
Adler, J., {\"O}ktem, O.: Learned primal-dual reconstruction. IEEE TMI  \textbf{37},  1322--1332 (2018)

\bibitem{cnngrad}
Harshit, G., et~al.: {CNN}-based projected gradient descent for consistent {CT} image reconstruction. IEEE TMI  \textbf{37},  1440--1453 (2018)

\bibitem{learn}
Chen, H., et~al.: {LEARN}: Learned experts' assessment-based reconstruction network for sparse-data {CT}. IEEE TMI  \textbf{37},  1333--1347 (2018)

\bibitem{admmct}
Wang, J., et~al.: {ADMM}-based deep reconstruction for limited-angle {CT}. Physics in Medicine \& Biology  \textbf{64} (2019)

\bibitem{dior}
Dianlin, H., et~al.: {DIOR}: Deep iterative optimization-based residual-learning for limited-angle {CT} reconstruction. IEEE TMI  \textbf{41},  1778--1790 (2022)

\bibitem{fistanet}
Jinxi, X., Yonggui, D., Yunjie, Y.: {FISTA}-{N}et: Learning a fast iterative shrinkage thresholding network for inverse problems in imaging. IEEE TMI  \textbf{40},  1329--1339 (2021)

\bibitem{regformer}
Xia, W., et~al.: Transformer-based iterative reconstruction model for sparse-view {CT} reconstruction. In: MICCAI. pp. 790--800 (2022)

\bibitem{aun}
Monga, V., Li, Y., Eldar, Y.C.: Algorithm unrolling: Interpretable, efficient deep learning for signal and image processing. IEEE Sig. Proc. Mag.  \textbf{38}(2),  18--44 (2021)

\bibitem{wdpm}
Phung, H., Dao, Q., Tran, A.: Wavelet diffusion models are fast and scalable image generators. In: CVPR. pp. 10199--10208 (2023)

\bibitem{wnerf}
Xu, M., et~al.: Wavenerf: Wavelet-based generalizable neural radiance fields. In: ICCV. pp. 18149--18158 (2023)

\bibitem{svd}
Rui, L., et~al.: Singular value decomposition-based {2D} image reconstruction for computed tomography. Journal of X-ray science and technology  \textbf{25},  113--134 (2017)

\bibitem{art}
Avinash, K., Malcolm, S.: Principles of Computerized Tomographic Imaging. Society for Industrial and Applied Mathematics (2001)

\bibitem{tv}
Tsutomu, G., Yukio, K.: Use of a {T}otal {V}ariation minimization iterative reconstruction algorithm to evaluate reduced projections during digital breast tomosynthesis. BioMed Research International pp. 1--14 (2018)

\bibitem{tikhonov}
Peng, C., Rodi, W.L., Toks\"{o}z, M.N.: A Tikhonov Regularization Method for Image Reconstruction. Springer US (1993)

\bibitem{learnpp}
Yi, Z., et~al.: {LEARN}++: Recurrent dual-domain reconstruction network for compressed sensing {CT}. IEEE TRPMS  \textbf{7},  132--142 (2023)

\bibitem{swint}
Liu, Z., et~al.: Swin transformer: Hierarchical vision transformer using shifted windows. In: ICCV. pp. 10012--10022 (2021)

\bibitem{aapm}
McCollough, C.: {TU-FG-207A-04}: Overview of the low dose {CT} grand challenge. Medical Physics  \textbf{43},  3759--3760 (2016)

\bibitem{adamw}
Loshchilov, I., Hutter, F.: {Decoupled Weight Decay Regularization}. arXiv preprint arXiv:1711.05101  (2017)

\end{thebibliography}
